\documentclass[lettersize,journal]{IEEEtran}

\IEEEoverridecommandlockouts

\usepackage{cite}
\usepackage{amsmath,amssymb,amsfonts}
\usepackage{graphicx}
\usepackage{algorithm}
\usepackage{textcomp}

\def\BibTeX{{\rm B\kern-.05em{\sc i\kern-.025em b}\kern-.08em
    T\kern-.1667em\lower.7ex\hbox{E}\kern-.125emX}}

\usepackage{boldline}
\usepackage{subfig}
\usepackage{bm}

\usepackage[noend]{algpseudocode}
\usepackage{algorithm}
\usepackage[hidelinks]{hyperref}
\usepackage[dvipsnames]{xcolor}

\algnewcommand{\LineComment}[1]{\State \(\triangleright\) #1}

\begin{document}

\title{Optimizing Multi-User Semantic Communication via Transfer Learning and Knowledge Distillation}

\author{Loc X. Nguyen, Kitae Kim,~\IEEEmembership{Member,~IEEE}, 
       Ye Lin Tun, Sheikh Salman Hassan,~\IEEEmembership{Member,~IEEE}, Yan~Kyaw~Tun,~\IEEEmembership{Member,~IEEE}, Zhu~Han,~\IEEEmembership{Fellow,~IEEE},~and~Choong~Seon~Hong,~\IEEEmembership{Fellow,~IEEE}
       
\thanks{Loc X. Nguyen, Kitae Kim, Ye Lin Tun, Sheikh Salman Hassan and Choong Seon Hong  are with the Department of Computer Science and Engineering, Kyung Hee University,  Yongin-si, Gyeonggi-do 17104, Rep. of Korea, e-mails:{\{xuanloc088, glideslope, yelintun, salman0335, cshong\}@khu.ac.kr}.}
\thanks{Yan Kyaw Tun is with the Department of Electronic Systems, Aalborg University, 2450 København SV, Denmark, e-mail:{\{ykt\}@es.aau.dk}.
}
\thanks{Zhu Han is with the Department of Electrical and Computer Engineering at the University of Houston, Houston, TX 77004 USA, and also with the Department of Computer Science and Engineering, Kyung Hee University, Seoul, South Korea, 446-701. email:{\{hanzhu22\}}@gmail.com.}       
}

\maketitle

\begin{abstract}

Semantic communication, notable for ensuring quality of service by jointly optimizing source and channel coding, effectively extracts data semantics, reduces transmission length, and mitigates channel noise. However, most studies overlook multi-user scenarios and resource availability, limiting real-world application. This paper addresses this gap by focusing on downlink communication from a base station to multiple users with varying computing capacities. Users employ variants of Swin transformer models for source decoding and a simple architecture for channel decoding. We propose a novel training regimen, incorporating transfer learning and knowledge distillation to improve low-computing users' performance. Extensive simulations validate the proposed methods.

\end{abstract}


\section{Introduction}

The exponential growth in the number of connected devices and the requirement for transmitting massive amounts of data for high-tech applications have raised several challenging problems for the conventional communication system. Conventional communication is built based on the accuracy of transmitting signals from the sender to the receiver, which is nearly approaching the Shannon physical-layer capacity limit \cite{luo2022semantic}, compelling researchers to imagine solutions for the future of wireless communication systems. Semantic communication has been considered as the pivotal technology to satisfy the quality of service (QoS) requirements for a substantial volume of devices. It is the second level of communication proposed by Shannon \cite{shannon1949mathematical}. The fundamental concept behind it is to eliminate unnecessary information within the message, and only transmit the semantics of the data. To achieve this, the researchers have considered jointly optimizing the source encoder and channel encoder, which not only embeds the meaning of the data into the transmission process but also improves the robustness in non-ideal channel environments.

The development of semantic communication systems has faced numerous challenges, primarily stemming from the need to embed the meaning of data into transmissions. Nonetheless, the proliferation of deep learning (DL) models has empowered semantic communication by facilitating the extraction of meaningful information from data and mitigating redundancy. Many existing works have proposed different DL-based joint source-channel coding (JSCC) to transmit text \cite{farsad2018deep,peng2022robust,hu2022one}, image \cite{10094735,bourtsoulatze2019deep}, and audio \cite{10447612,weng2021semantic}. These studies focus on performance improvement by adopting the latest developments in the DL architecture, such as Convolutional Neural Network (CNN) and Transformer. Only a few examined multi-user scenarios in the context of semantic communication \cite{xie2022task,9814491}, while none of the aforementioned works considered the difference in users' resources, such as storage and computing capacity. In \cite{10423076}, the authors considered the difference in the computing capacity of users. Nevertheless, the iterative training process proposed herein necessitates a substantial number of training epochs to achieve convergence, resulting in resource-intensive and time-consuming operations.

In this paper, we consider the image transmission from the base station (BS) to multiple users with different computing capacities (equipped with variants of Swin transformer) in the semantic communication system. This problem can be challenging due to the difference in user decoders, while only one encoder is available at the BS. On the other hand, deploying multiple encoders at the BS to serve multi-users can be unscalable in terms of storage. Therefore, we propose a training procedure to stabilize the JSCC at the BS while maintaining good performance. Finally, we present two techniques to significantly boost the low-computing user's performance: transfer learning and knowledge distillation \cite{44873}. The main contributions of our paper are summarized as follows:
\begin{figure*}[t!]
    \centering
    \includegraphics[width=0.765\textwidth]{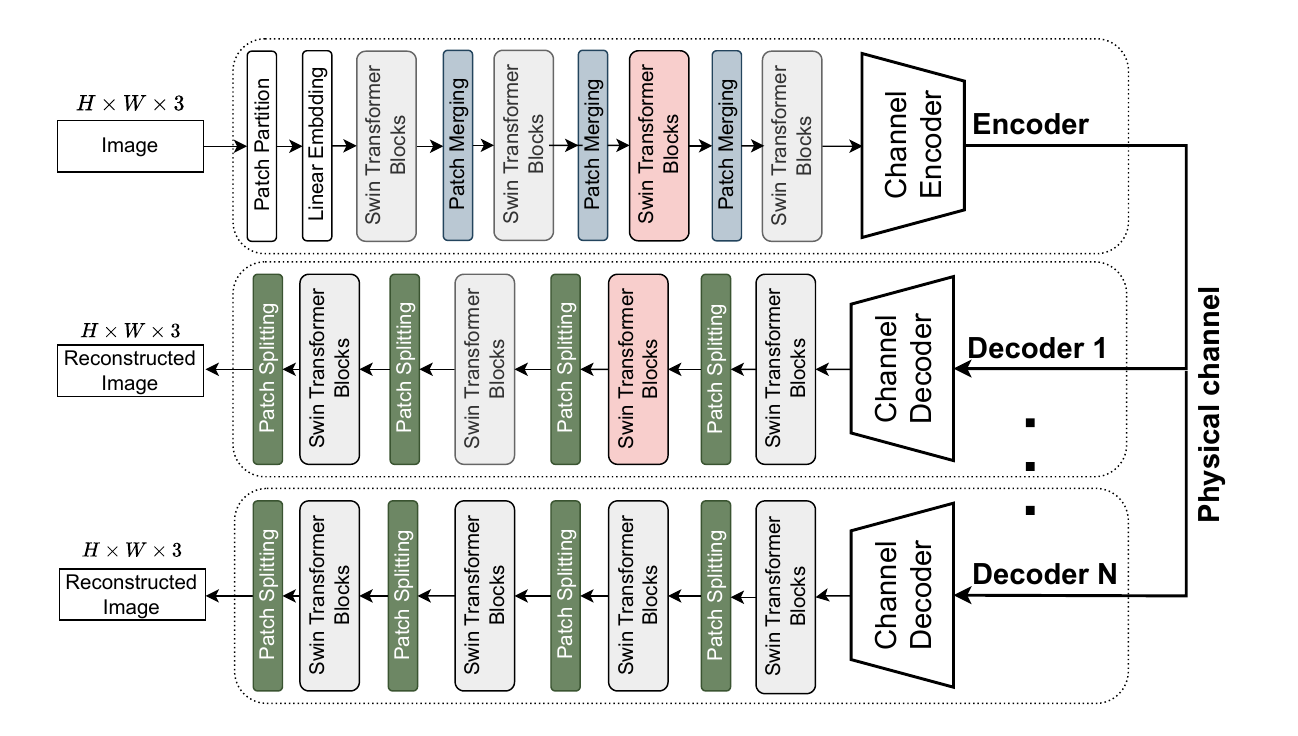}
    \caption{The system model includes one transmitter and two receivers. The difference in the number of Swin-Transformer blocks leads to a difference in computing capacities between receivers.}
    \label{systemmodel}
\end{figure*}
\begin{itemize}
    \item We consider image transmission from the BS to users with varying computing capacities in the network. To address potential inefficiencies when training with different decoders, we first pair the joint source-channel coder (JSCC) at the BS with one user and train this pair to achieve optimal performance. Then, we freeze the encoder and train the other users to adapt to the frozen encoder. This procedure not only stabilizes but also accelerates the training process. 
    \item We transfer parameters from the trained user to the untrained user for identical layers, which improves the untrained decoder's performance and training efficiency. 
    \item In addition, the knowledge distillation technique is leveraged in our system. Specifically, the trained user acts as the teacher, while the untrained user is the student. The technique demonstrates outstanding results in the quality of the reconstruction image. 
    \item Finally, we conduct an extensive series of simulations to prove the efficiency of the training procedure and the two advanced techniques proposed in our semantic communication system for multi-users. Specifically, the proposed training procedure achieves higher metric values in every signal-to-noise ratio (SNR) for the low-computing user.
\end{itemize}

\section{System Model}

As shown in Fig.~\ref{systemmodel}, we consider the downlink transmission from the BS to multi-users in semantic communication, particularly focusing on image reconstruction. Unlike a conventional communication system, semantic communication employs the joint source-channel encoding procedure to consider the semantic meaning of the data in the transmission process. For notation simplification, we designate `decoder 1' and `decoder 2' to represent the decoders affiliated with users 1 and 2, respectively. With a given input image $I$, the BS first extracts its semantic features with the semantic encoder and then compresses those features with the channel encoder. This process is presented as: 
\begin{equation}
    X_{I}= E^{C}_{\beta}(E^{S}_{\psi}(I)),
\end{equation}
where $E^{C}_{\beta}$ and $E^{S}_{\psi}$, respectively, denote the channel encoder and semantic encoder, while $\beta$ and $\psi$ are the network parameters. $X_{I}$ is the output of the channel encoder, which will be transmitted to the user through a wireless environment. We mainly examine the additive white Gaussian noise (AWGN) channel as the physical communication channel, while the channel fading effect can be minimized with the channel state information (CSI) as in \cite{xie2021task}. Therefore, the received signal at the user $k$ is expressed as follows:
\begin{equation}
    Y_{I,k}= X_{I}+N_{k},
\end{equation}
where $N_{k}$ denotes the AWGN from the wireless environment from the BS to receiver $k$, whose elements follow the  $\mathcal{CN}(0,\sigma^{2}\boldsymbol{I})$ distribution. User $k$ inputs the received signal into its joint source channel decoder as: 
\begin{equation}
    \hat{I}_{k}= D^{S}_{\sigma_{k}}(D^{C}_{\gamma_{k}}(Y_{I,k})),
\end{equation}
where $D^{S}_{\sigma_{k}}$ and $D^{C}_{\gamma_{k}}$ denote the source and channel decoder of the user $k$, while $\sigma_{k}$ and $\gamma_{k}$ are the corresponding parameters. With the reconstructed image $\hat{I}_{k}$, the most common loss function to train the whole network is the mean square error.
\begin{equation}\label{MSEloss}
    \mathcal{L}(I,\hat{I}_{k})= \mathrm{MSE}(I,\hat{I}_{k}).
\end{equation}

\section{Multi-users in Semantic Communication}

We consider two users with different computing capacities. The architectural complexity of the source encoder and decoder surpasses that of the channel coder, as their primary task involves extracting the semantic content of the data, in contrast to the feature compression performed by the latter. Therefore, this paper considers the computational differences in the
source decoder. Specifically, both the high-computing and the low-computing users are equipped with four stages of Swin transformers. The high-computing decoder has a stage block configuration of (2, 6, 2, 2), while the low-computing user has only two blocks per stage in the source decoder. Meanwhile, the source encoder also has four stages of Swin transformers with a stage block configuration of (2, 2, 6, 2).

The most straightforward approach to training this network involves pairing the encoder with one decoder for one epoch, then switching to train with the other decoder in the subsequent epoch, as described in \cite{10423076}. However, this method requires a considerable number of epochs to achieve convergence, leading to prolonged training times and significant resource demands. To address these challenges, we propose a novel approach to network training, along with two techniques to expedite the training process.

\subsection{Proposed Training Procedure}
Training two decoders iteratively at each epoch can confuse the encoder regarding its convergence direction. Therefore, we first couple the encoder with the high-computing decoder and train them together for several epochs. Training one decoder at a time provides a clear direction for the encoder to extract and compress meaningful messages effectively. Afterward, we freeze the encoder and train the other decoder to reconstruct the image based on the received signal from the trained encoder.
\subsection{Partially Transfer Learning between Users}
Recognizing the identical architecture in the last two blocks of the source decoders, we transfer the parameters from the trained high-computing decoder to the low-computing decoder to study the effect on the training process. This technique aims to provide a better set of parameters than random initialization. Inspired by transfer learning, this approach can be considered a form of it. When training the entire low-computing decoder network, the transferred parameters significantly guide the network to converge effectively. Our simulation results also present the performance when the transferred parameters are frozen, and the remaining parameters are fine-tuned.
\subsection{Knowledge Distillation from the Teacher to the Student}
The partial weight transfer mechanism only works under the assumption that two decoders share identical structures in some layers, which is not always the case. Here, we consider the second approach to provide an efficient training process: knowledge distillation. Instead of training the low-computing user by only using the MSE between the reconstruction and original images, we include the MSE between the reconstructed images of two decoders into the training process as knowledge distillation from the teacher (trained high-computing user) to the student (untrained low-computing user). The training loss is described as:
\begin{subequations}
\begin{align}
\mathcal{L}_{\mathrm{distill}}= \mathrm{MSE}(\hat{I}_1, \hat{I}_2), \label{distillloss} \\ 
 \mathcal{L}_{\mathrm{train}}=  \mathcal{L}(I,\hat{I}_2) + \alpha\mathcal{L}_{\mathrm{distill}},  \label{combinedloss}
\end{align}
\end{subequations}
where $\alpha$ denotes the weight of the distillation loss. The concept behind knowledge distillation is to mimic the outputs of the teacher model rather than directly optimizing for the reconstruction task at hand. This phenomenon particularly fits our system for the following reasons: it is impossible to completely eliminate all the noise from the physical environment and reconstruct the image exactly the same as the original image. Therefore, mimicking the output of the trained high-computing decoder (teacher) can guide the low-computing decoder (student) to get better performance instead of finding an absolute answer. The training details can be found in Algorithm~\ref{alg:Alg1}.

\begin{algorithm}[t]
   \caption{\strut Multi-user Semantic Communication with Knowledge Distillation} 
   \label{alg:Alg1}
   \begin{algorithmic}[1]
       \State{\textbf{Initialize:} Encoder at the BS $E^{C}, E^{S}$, high-computing decoder $D^C_1,D^S_1$, low-computing decoder $D^C_2,D^S_2$, training epochs for each decoder $T$.}
        \State{\textbf{Training high-computing decoder}}
        \For{each epoch t$=1,2,...,T$}
          \For{each image batch}
                \State{Sampling the noise $N_{1}$ from the set of SNR.} 
                \State{$X_{I}= E^{C}_{\beta}(E^{S}_{\psi}(I))$}
                \State{$Y_{I,1}= X_{I}+N_{1},$}
                \State{$\hat{I}_{1}= D^{S}_{\sigma_{1}}(D^{C}_{\gamma_{1}}(Y_{I,1})),$}
                \State{$\mathcal{L}(I,\hat{I}_{k})= \mathrm{MSE}(I,\hat{I}_{k})$}
                \State{Update $E^{C}, E^{S}, D^C_1,D^S_1$ with $\mathcal{L}(I,\hat{I}_{k})$.}
          \EndFor
                \EndFor
    \State{\textbf{Output:} Trained encoder and high-computing decoder.}
        \State{Freeze $E^{C}, E^{S}, D^C_1,D^S_1$.}
        \State{\textbf{Training low-computing decoder with KD}}
        \For{each epoch t$=1,2,...,T$}
          \For{each image batch}
                \State{Sampling the noise $N_{2}$ from the set of SNR.} 
                \State{$X_{I}= E^{C}_{\beta}(E^{S}_{\psi}(I))$}
                \State{$Y_{I,2}= X_{I}+N_{2},.$}
                \State{$\hat{I}_{2}= D^{S}_{\sigma_{2}}(D^{C}_{\gamma_{2}}(Y_{I,2})),$}
                \State{$\hat{I}_{1}= D^{S}_{\sigma_{1}}(D^{C}_{\gamma_{1}}(Y_{I,2})),$}
                \State{$\mathcal{L}_{\mathrm{train}}=  \mathcal{L}(I,\hat{I}_2) + \alpha\mathcal{L}_{\mathrm{distill}}$}
                \State{Update $D^C_2,D^S_2$ with$\mathcal{L}_{\mathrm{train}}$}
          \EndFor
        \EndFor
    \State{\textbf{Output:} Trained semantic communication network for multi-users.}
   \end{algorithmic}
\end{algorithm}

\section{Simulation Results}

In this section, we carry out a series of simulation scenarios to analyze the efficiency of the proposed system for multi-user training. The parameter setting for the scenario is as follows: \textit{Compression ratio} $1/16$, \textit{learning rate} $5e^{-4}$, \textit{batch size} $8$, \textit{training epochs} $500$,  \textit{the range of SNR value} $1, 3, 5, 7$.

\subsection{Model and Dataset}

\emph{Joint source-channel encoder/decoder model}: We leverage the Swin Transformer model as the source encoder/decoder to extract the semantics of the data while the channel encoder/decoder is built based on the seven connected layers and a skip connection as \cite{10094735}. We use the DIV2K \cite{agustsson2017ntire} training dataset to train the system, which includes eight hundred 2K resolution images. On the other hand, the DIV2K testing dataset is utilized to evaluate the trained network, and the provided results are the average value of all the testing images. 

\subsection{Benchmark}
To thoroughly assess the effectiveness of our training procedure alongside the other two DL techniques, we consider the following comprehensive benchmarks.

\begin{itemize}
    \item \emph{Iterative training:} The joint source-channel encoder at the BS is iteratively coupled and trained with the decoders of varying capacity.
    \item \emph{The proposed training procedure:} The encoder couples and trains with the high-computing decoder, and then we freeze its parameters to train the low-computing decoder.
    \item \emph{The proposed training procedure \& Partial transfer learning:} The low-computing decoder partially received the parameters of the high-computing.
    \item \emph{The proposed training procedure \& Knowledge distillation:} The low-computing decoder is trained to reconstruct the original and mimic the high-computing decoder.
\end{itemize}

\begin{figure}[t!]
	\centering
	\includegraphics[width=0.85\linewidth]{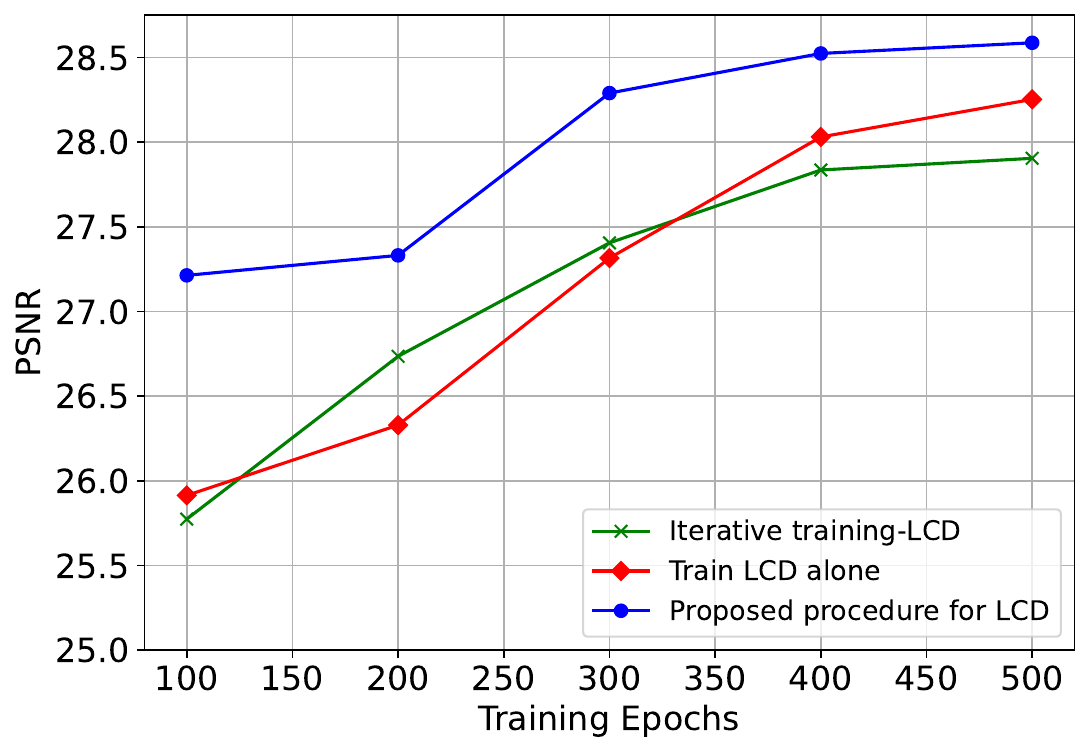}
	\caption{\textcolor{black}{The PSNR values of the proposed procedure, iterative training, and the train alone under SNR = 3dB.}}
        \label{ConvergenceSpeed}
\end{figure}
\begin{figure}[t!]
	\centering
	\includegraphics[width=0.85\linewidth]{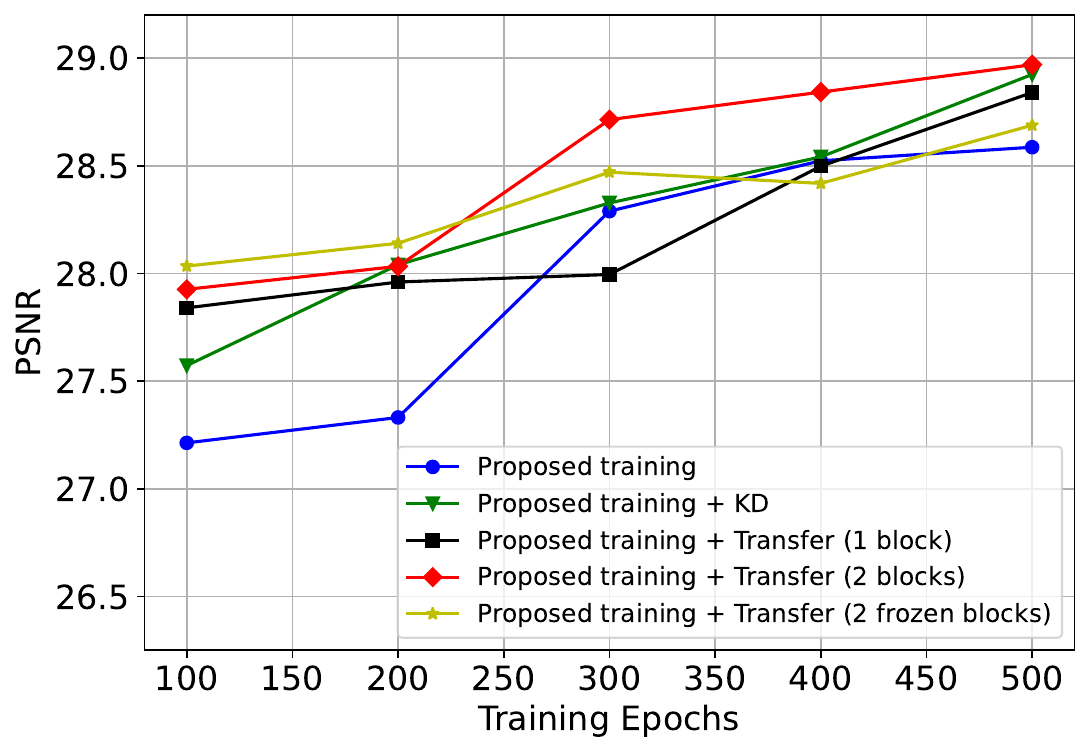}
	\caption{The effects of transfer learning and knowledge distillation on the performance of LCD under SNR = 3dB.}
        \label{transferanddistill}
\end{figure}

\subsection{Evaluating Convergence Speed of the Proposed Procedure for Low-computing Decoder}
In this sub-section, we evaluate the performance of the system under different training procedures: train alone, train iteratively, and the proposed method. It is important to note that in the iterative training case, the total number of epochs is 1000. When the training epoch of the low-computing decoder (LCD) is 200, it means the encoder has been trained for 400 epochs by both decoders. Fig.~\ref{ConvergenceSpeed} shows the performance of the LCD using the Peak Signal to Noise Ratio (PSNR) metric \cite{hore2010image}, which is widely used to assess the quality of reconstructed images. Since the proposed training procedure only affects LCD training, we exclude the performance of the high-computing decoder (HCD) to simplify the figure. The proposed procedure for training the LCD achieves a faster convergence rate and higher performance compared to the iterative training procedure and the train-alone case. This result can be explained as follows: 1) During the first stage, the coupling between the encoder and the HCD for training enables the encoder to focus on a single objective, successfully extracting the semantic features of the data. 2) In the second stage, with the correct semantic features extracted, the training process of the LCD becomes much more stable due to the reduced number of parameters being updated (as the encoder parameters are frozen).
\begin{figure}[t!]
	\centering
	\includegraphics[width=0.85\linewidth]{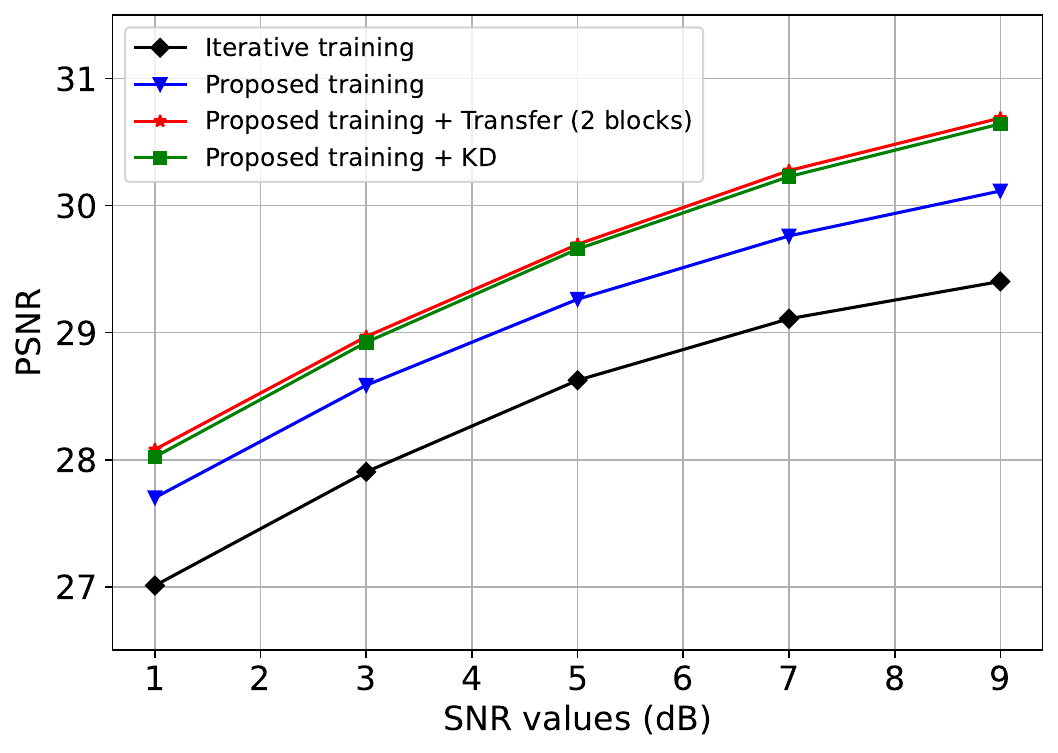}
	\caption{The PSNR values of LCD after 500 epochs under different channel conditions.}
        \label{Robust}
\end{figure}
\subsection{Partially Transfer Learning and Knowledge Distillation Evaluation}
In this sub-section, we apply either partial transfer learning or knowledge distillation on top of our proposed training process, utilizing the parameters and knowledge of the HCD. Specifically, we transfer the last two blocks of the source decoder from the HCD to the LCD, as they share an identical network architecture. As shown in Fig.~\ref{transferanddistill}, the transfer learning technique achieves the highest performance by providing the LCD with a subset of parameters to initiate the training process. It performs even better than our proposed training procedure when the transferred parameters are frozen. The more parameters we transfer, the higher the network's performance. However, a drawback of transfer learning is that it requires an identical network architecture in some layers, unlike the case for the knowledge distillation technique. The LCD trained with knowledge distillation steadily improves its performance and achieves relatively good results. It outperforms the normal proposed procedure but falls slightly short of the transfer learning technique.
\begin{figure*}[t]
    \centering
    \includegraphics[width=0.9\textwidth]{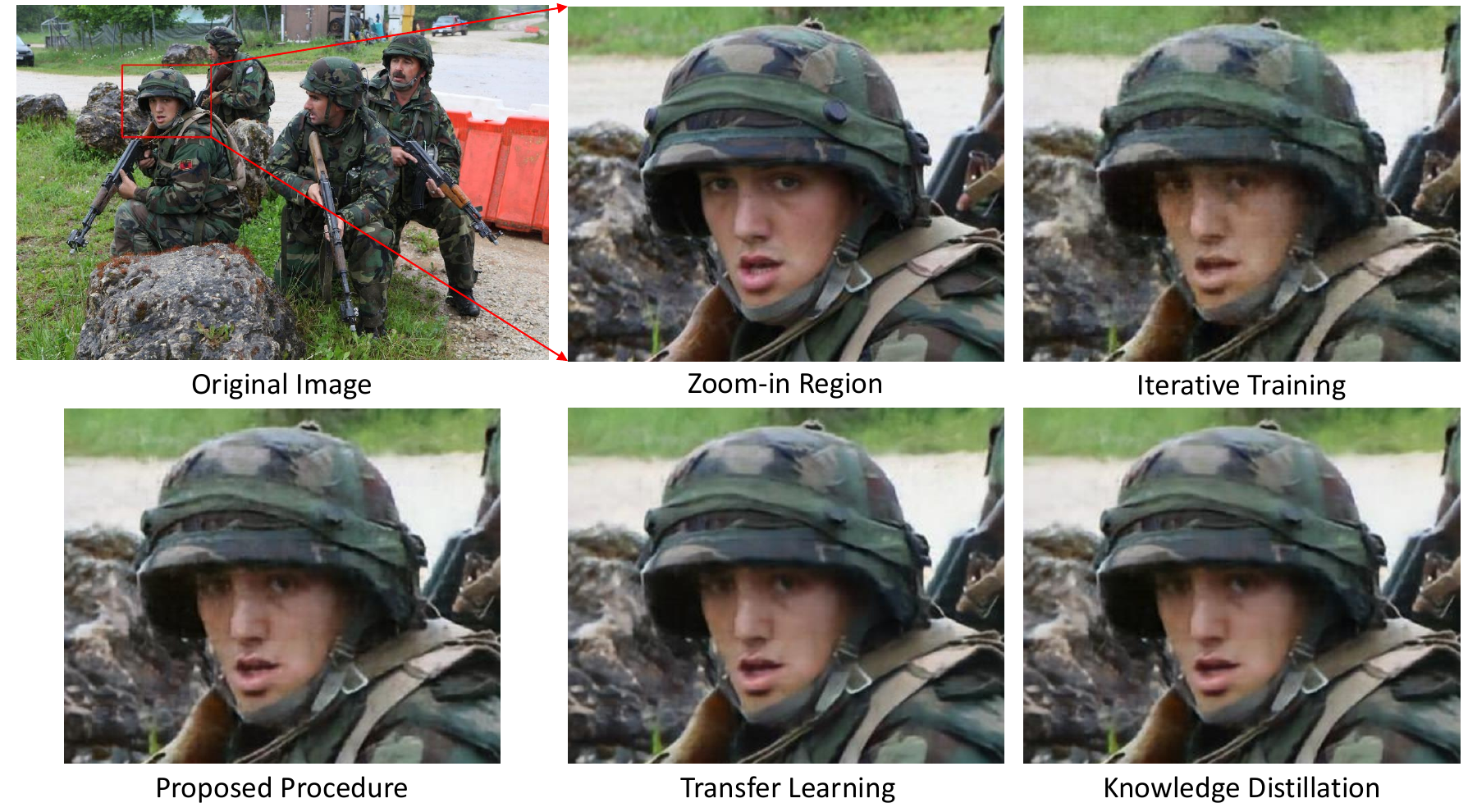}
    \caption{The image reconstructed by different training methods under the SNR = 3dB.}
    \label{ImageVisual}
\end{figure*}
Fig.~\ref{Robust} provides the performance of the LCD when the channel noise varies, in which a higher value of the SNR indicates a lower magnitude of noise in the environment. Therefore, we can see the performance of all the training scenarios increases when the SNR value increases. The transfer learning and knowledge distillation techniques achieve similar results in all cases and surpass the other two schemes by a large margin. In addition, we provide the reconstructed images of these schemes in Fig.~\ref{ImageVisual}. It is easy to spot the noise in the human face of the reconstructed image by iterative training, while the proposed procedure generates a smoother image. Similarly, the visual quality of the images produced by transfer learning and knowledge distillation is smooth and closely resembles the original images.

\section{Conclusion}

In this paper, we have studied a multi-user semantic communication system with users having different computing capacities, reflecting real-world scenarios. We have identified the issue of slow convergence when iteratively training the joint source-channel encoder at the base station with different decoders to be deployed in the network. To address this, we have proposed a new training procedure that stabilizes and accelerates the convergence of the training process. By initially training the encoder with the highest-capacity decoder, our method is scalable in a way that an arbitrary number of decoders with different capacities can be trained afterward. This can support users with varying resources in the network. Additionally, we have adopted two advanced deep-learning techniques, transfer learning and knowledge distillation, to enhance the performance of low-computing decoders. Our simulation results demonstrate the effectiveness of our training procedure, significantly outperforming the iterative training approach. The parameter transfer provides the low-computing decoder an initial advantage, while the knowledge distillation guides it to a feasible and effective solution.

\bibliographystyle{IEEEtran}
\bibliography{main}

\end{document}